\documentstyle[prb,aps,preprint,tighten]{revtex}

\begin{document}

\draft

\title{Generalized Bloch equations for a strongly driven tunneling system}

\author{Peter Neu$^{1)}$ and   Jochen Rau$^{2)}$\footnote{Present
address: Max-Planck-Institut f\"ur Physik komplexer Systeme,
Bayreuther Stra{\ss}e 40 Haus 16, 01187 Dresden, Germany.}}

\date{November 7, 1996}

\address{$^{1)}$Department of Chemistry,  
 Massachusetts Institute of Technology, Cambridge MA
02139, USA\\
$^{2)}$European Centre for Theoretical Studies in Nuclear Physics
and Related Areas (ECT$^\star$), Villa Tambosi, Strada delle Tabarelle 286,
38050 Villazzano (Trento), Italy}

\maketitle
\thispagestyle{empty}

\vspace{6cm}
\begin{abstract}   
Using the {\it Robertson projection operator formalism},
we derive generalized Bloch equations which describe
the dynamics of a biased two-level tunneling system
strongly driven
by an external field
and weakly coupled to a super-Ohmic
heat bath.
The generalized Bloch equations constitute a set of coupled
nonlinear integro-differential equations.
With their help we investigate
the influence of phonons on the phenomenon of dynamical localization.
\end{abstract}



\newpage

\setcounter{page}{1}

\section{Introduction}

The last decade has witnessed detailed investigations of the problem of
macroscopic quantum coherence.\cite{Legg,Weisb}
The generic model is a two-level tunneling
system (TLS) coupled to a harmonic heat bath and an external driving field.
The tunneling units may represent, e. g.,  small groups of atoms or molecules
in structurally disordered solids, such as in the study of low-temperature
properties of glasses\cite{P};  electrons  in semiconductor
quantum wells\cite{Esaki} or in the electron-transfer dynamics
in chemical reactions\cite{FW}; or  magnetic flux quanta at  
Josephson-junctions or  SQUIDs\cite{Le}.
The basic mechanism by which quantum coherence is  destroyed
is the coupling of the quantum system to the dissipative environment.
Various approaches using
second-order perturbation theory, variational polaron theory\cite{Bob1},
mode-coupling theories\cite{mode} or, most popular, the 
noninteracting-blip approximation (NIBA)\cite{Legg,Weisb}
have been used to discuss the decoherence effect of the bath.

One key issue is whether by careful tuning of the driving field
the effect of dissipation may be reduced and quantum coherence 
be restored on the macroscopic level.
 This idea of controlling
tunneling and relaxation has become especially popular
since the discovery, by Grossmann, H\"anggi and coworkers\cite{Haeng},
of the effect of {\it coherent destruction of tunneling}.
In a numerical treatment of an isolated bistable quartic potential
driven by a periodic force
-- but not coupled to a heat bath --
they found 
complete suppression  of tunneling: 
at certain ratios of the field amplitude 
and the driving frequency the two lowest Floquet eigenstates (quasienergies)
of the driven system become degenerate, thereby preventing
the system from performing coherent oscillations between 
the left and right well. As a result, 
a particle initially localized in one well
will remain there forever. This so-called
{\em dynamical localization} effect has also been described in a 
two-state model,\cite{Haeng2,add,DM} and
has since been the subject of several
further investigations.\cite{Ma,Milena,Ditt,Yu1,Yu2,Maki,BM,CP,Mak,NS}

Recent studies have focused on the question to what extent
the dynamical
localization of a TLS will be affected by
the coupling to a heat bath. 
Physical intuition seems to suggest that in order for 
dynamical localization to occur, 
a wave paket initially localized in one well must have a
well-defined phase; only then can one expect that a mismatch between
the tunneling motion and the driving field may prevent a particle
from escaping into the other well. Since phonons destroy this
phase coherence -- i. e., there is a finite phase memory time
$\tau_2$ --
 dynamical localization should always be softened 
through the coupling to a heat bath.  

Recently 
Grifoni {\it et al.}\cite{Ma} gave
a systematic approach to the transient and steady-state dynamics 
of a driven dissipative TLS.
They
considered Ohmic and frequency-dependent damping  in the framework
of the NIBA for the stochastic force, but without any
approximation for the driving force. In Ref. \CITE{Milena} they 
extended their approach beyond the NIBA.
Dittrich {\it et al.}\cite{Ditt}
investigated numerically the effect of driving and dissipation 
on the coherent tunneling motion of a symmetric bistable system for 
weak Ohmic damping.
Dakhnovskii\cite{Yu1,Yu2} employed small-polaron theory to
address the same issue in a two-state approximation within the NIBA. 
He concluded for the case of a symmetric TLS 
with super-Ohmic damping
that, in contrast to intuition, the localization
transition  remains stable against  disturbance by the bath
as long as the driving frequency is larger than the relaxation energy (polaron-binding
energy)  of the lattice.
On these grounds he predicted
the existence of a slow mode oscillation near the transition.\cite{Yu1}
 His conclusions essentially rely on the fact that near the localization
transition the rate of phase loss -- which is due to the bath --
decreases faster than the tunneling coherence.

So far most
investigations have made use of the NIBA. 
Yet the NIBA breaks down at non-zero bias, weak
dissipation and low temperatures.
It is the latter regime, so far poorly understood, which we
shall discuss in this paper.
Weak TLS-phonon coupling is the limit which is relevant for, e. g.,
the description of tunneling
 centers in dielectric solids.
We will work within the two-state picture and assume a super-Ohmic
spectral density for the phonons; 
and we will treat
the TLS-phonon coupling
in Born approximation.\cite{footnote_born}
(There are in fact general arguments why
under the above assumptions it should always be justified to treat the dissipation 
perturbatively.\cite{Legg})
The external field, on the other hand, will be treated to any order. 
 We are then able to derive nonlinear and non-Markovian
  {\it generalized Bloch equations} (GBE).
These in turn permit us to discuss the influence of weak dissipation
on the localization transition for both the symmetric and the biased case, and
for arbitrarily strong driving and any temperature.
In the high-temperature and high-frequency limit there are analytical solutions,
which we will compare with those obtained from
the NIBA.

The paper is organized as follows. In Sec. II we 
give a short introduction to the Robertson formalism and apply it
to driven TLS-dynamics. In Sec. III we derive the GBE.
The linear response regime is discussed as a special case in Sec. IV.
In Sec. V we address the influence 
of weak dissipation on the dynamical localization transition,
before we close (in Sec. VI) with a brief summary.

\setcounter{equation}{0}

\section{Robertson Formalism}
We will derive a non-Markovian generalization of the Bloch equations
with the help of the so-called Robertson formalism
\cite{Robert}.
The Robertson formalism 
constitutes a particular variant
of the well-known projection technique
\cite{transport}, one that is tailored to 
describing the evolution of selected
observables
even far from equilibrium.

When studying the dynamics of a macroscopic quantum system,
one is typically confronted with the problem of determining
the evolution of only a small set of selected (``relevant'')
observables $\{G_a\}$.
Let the (generally time-dependent) Hamiltonian be denoted
by $H(t)$.
Then the equation of motion for the expectation values
\begin{equation}\label{exvalues}
g_a(t)\equiv\langle G_a\rangle_{\rho(t)}:=
\mbox{tr}[\rho(t)G_a]
\end{equation}
reads
\begin{equation}\label{lvn}
\dot g_a(t)=i\langle{\cal L}(t)G_a\rangle_{\rho(t)}
\quad,
\end{equation}
with $\rho(t)$ being the statistical operator and
${\cal L}(t)$ the Liouvillian
\begin{equation}
{\cal L}(t):=\hbar^{-1}[H(t),*]
\quad.
\end{equation}
Associated with the
expectation values $\{g_a(t)\}$ is, at each time $t$,
a generalized canonical state
\begin{equation}\label{gen_canon}
\rho_{\rm rel}(t):=Z(t)^{-1}\exp\left(
-\sum_a l^a(t) G_a\right)
\quad,
\end{equation}
called the {\em relevant part of the statistical
operator}, with partition function
\begin{equation}
Z(t):=\mbox{tr}\,\exp\left(
-\sum_a l^a(t) G_a\right)
\end{equation}
and the Lagrange parameters $\{l^a(t)\}$ adjusted such
as to yield the correct expectation values (\ref{exvalues}) of the
relevant observables.
The difference
\begin{equation}
\rho_{\rm irr}(t):=\rho(t)-\rho_{\rm rel}(t)
\end{equation}
is then the {\em irrelevant part} of the state.

Like all variants of the projection technique,
the Robertson formalism is
based on a clever insertion of projection operators
into the equation of motion (\ref{lvn}).
Here the projection operator ${\cal P}(t)$ is chosen such that it projects
arbitrary vectors in Liouville space onto the
subspace spanned by the unit operator and by the 
relevant observables $\{G_a\}$, and that this projection
is orthogonal with respect to the (time-dependent) scalar product
\begin{equation}
\langle A;B\rangle^{(t)} :=
\int_0^1 d\mu\,\mbox{tr}\left[\rho_{\rm rel}(t)^\mu
A^\dagger \rho_{\rm rel}(t)^{1-\mu}B\right]
\quad.
\end{equation}
Since this scalar product varies in time, the projection
operator, too, is time-dependent;
so is its complement
${\cal Q}(t) :=1-{\cal P}(t)$.
The projection operator is known as 
the {\em Kawasaki-Gunton projector}.\cite{kawasaki}

Let
${\cal T}(t',t)$ be the evolution operator defined by
\begin{equation}
\label{irrel:evolution}
{\partial\over\partial t'}{\cal T}(t',t)=
-{i}\,{\cal Q}(t'){\cal L}(t'){\cal Q}(t')
{\cal T}(t',t) 
\end{equation}
with initial condition
${\cal T}(t,t)=1$.
As ${\cal Q}$ projects out the irrelevant component
of an observable, this operator
may be pictured as describing
the evolution of the {\em irrelevant} degrees of freedom
of the system.
The equation of motion
for the expectation values of the relevant observables
can then be cast into the form
\begin{eqnarray}
\dot g_a (t)
&=&
{i}\langle{\cal L}(t)G_a\rangle_{{\rm rel}(t)}
+\int_{0}^t\mbox{d}t'\sum_c
\langle{\cal Q}(t'){\cal L}(t')G_c;{\cal T}(t',t)
{\cal Q}(t){\cal L}(t)G_a
\rangle^{(t')}l^c(t')
\nonumber \\
&&
+{i}\langle{\cal T}(0,t){\cal Q}(t){\cal L}(t)G_a
\rangle_{{\rm irr}(0)}
\quad,
\label{robertson_2}
\end{eqnarray}
which is known as the
{\em Robertson equation}.
\cite{Robert}
Here $\langle\cdot\rangle_{\rm rel}$ and
$\langle\cdot\rangle_{\rm irr}$ denote expectation values
evaluated in the relevant and irrelevant part of the state,
respectively.
A major simplification occurs if, as is often the case,
the initial macrostate can be characterized completely
by the expectation
values $\{\langle{G_a}\rangle(0)\}$ of the relevant observables
and hence has the 
generalized canonical form (\ref{gen_canon}).
In this case the initial state has no irrelevant component,
which in turn implies that the third (``residual force'') term vanishes.
 
The Robertson equation embodies
a closed system of nonlinear coupled integro-differential equations
for the expectation values 
$\{g_a(t)\}$.
These coupled equations are non-local in time:
future expectation values of the relevant observables are
predicted not just on the basis of their present values,
but on their entire history.
In essence, the irrelevant degrees of freedom 
have been eliminated,
but this elimination must be paid for by
nonlinear and non-Markovian features of the projected
equation of motion;
the complexity of the evolution 
has been mapped onto nonlinearities and a non-local
behavior in time.

We now determine the Robertson equation for a driven TLS.
The underlying microscopic dynamics of the TLS is determined
by a time-dependent
Hamiltonian
\begin{equation}
H(t)=H_{\rm S}(t)+H_{\rm SB} + H_{\rm B}
\quad.
\end{equation}
Here the system Hamiltonian is given by
\begin{equation} \label{HS}
H_{\rm S}(t)=-{\textstyle{1\over2}}\vec{\sigma}\cdot\vec{h}(t)
\quad,
\end{equation}
where
$\vec{h}(t)=\hbar\cdot(\Delta_0,0,-\Delta-f(t))$
includes the tunneling frequency $\Delta_0$, the bias $\Delta$,
and the time-dependent driving field
\begin{equation}
f(t)=2\Omega\cos(\omega_L t)
\end{equation}
with Rabi frequency $\Omega$ and driving frequency $\omega_L$.
If we denote the state of a particle localized in the left (right)
well by $|L\rangle$ ($|R\rangle$) then the Pauli matrix
$\sigma_z=|L\rangle\langle L|-|R\rangle\langle R|$ represents its
coordinate, while $\sigma_x=|L\rangle\langle R|+|R\rangle\langle L|$
represents its tunneling motion.
With this convention $\Delta>0$ means that the left well has a 
higher potential energy than the right well.
The bath Hamiltonian
$H_{\rm B}=\sum_k \hbar\omega_k b^\dagger_k b_k$,
with bosonic creation and annihilation operators satisfying
$[b_i,b_j^\dagger]=\delta_{ij}$,
describes the dynamics of an ensemble of harmonic oscillators (phonons).
Their spectral density is taken to be super-Ohmic,
\begin{equation}
J(\omega)={2\over\hbar^2}\sum_j c_j^2 \delta(\omega-\omega_j)
=U\omega^3\exp(-\omega/\omega_c)
\quad,\quad \omega\ge 0
\quad,
\end{equation}
up to some cutoff $\omega_c$.
Finally, the system-bath coupling is assumed to be bilinear
in the tunneling and the phonon coordinate,
\begin{equation}
H_{\rm SB}=\sigma_z\sum_j c_j(b_j+b_j^\dagger)=:\sigma_z e
\quad.
\end{equation}
 
Relevant observables are
\begin{equation}
\{G_a\}\to\{{\sigma_\alpha},(H_{\rm SB}+H_{\rm B})\}
\quad,
\end{equation}
with expectation values 
\begin{equation}
\{g_a(t)\}\to\{{p_\alpha}(t),
E_{\rm B}(t)\}
\end{equation}
and associated Lagrange parameters
\begin{equation}
\{l^a(t)\}\to\{-\beta{\lambda_\alpha}(t)/2, \beta\}
\quad.
\end{equation}
Here $p_\alpha(t)$ denotes the time-dependent polarization
of the TLS, and $E_{\rm B}(t)$ the internal energy of the bath.
We assume the heat bath to be large enough so that
its inverse temperature $\beta=1/k_B T$ does not vary in time.
The relevant part of the statistical operator
reads
\begin{eqnarray}
\rho_{\rm rel}(t)
&=&
Z^{-1}\exp\left[-\beta\left(H_{\rm B}+H_{\rm SB}
-\textstyle{1\over2}\vec{\lambda}(t)\cdot\vec{\sigma}
\right)\right]
\nonumber \\
&=&
Z^{-1}\exp\left[-\beta\left(H(t)+
\textstyle{1\over2}(\vec{h}(t)-\vec{\lambda}(t))
\cdot\vec{\sigma}
\right)\right]
\quad.
\end{eqnarray}
Provided the (full) initial state $\rho(0)$ 
can be written
in this generalized canonical form, i. e., provided
$\rho(0)=\rho_{\rm rel}(0)$,
then the residual force term in the Robertson equation
vanishes. 
The class of initial states in the generalized canonical form
includes the common case of
a particle initially held at the site $|L\rangle$,
$p_z(0) = 1$, and coupled to a bath in thermal equilibrium,
\begin{equation}
\rho(0)\, \propto\,  \pi_{\rm L}\, \exp\left[-\beta\left(H(0)+
\textstyle{1\over2}\hbar\Delta_0\sigma_x
\right)\right]\, \pi_{\rm L}
\quad.
\end{equation}
Here, $\pi_{\rm L} = {1\over 2} ({\bf 1} + \sigma_z)$
denotes the projector on the state $|L\rangle$.
This initial condition has the generalized canonical form
(\ref{gen_canon}) with Lagrange parameters
$\vec{\lambda}(0) = (0,0,\infty)$.
Clearly, the fact that the initial state has the generalized
canonical form does not imply that the total system is in
equilibrium; rather, it means that the initial state can be
completely characterized by the expectation values 
$\{p_\alpha(0),E_B(0)\}$ of the
relevant observables.
The Robertson equation then translates
into the generalized Bloch equations (GBE)
\begin{equation}
\dot p_\alpha(t)=
\hbar^{-1}[\vec{p}(t)\times(\vec{h}(t)-\vec{\lambda}(t))]_\alpha
-\int_0^t dt'\sum_\beta K_{\alpha\beta}(t,t')
[\lambda_\beta(t')-h_\beta(t')]
\end{equation}
with memory kernel
\begin{equation}
K_{\alpha\beta}(t,t'):=
{\textstyle {\beta\over2}}
\langle {\cal Q}(t'){\cal L}_{\rm SB}\sigma_\beta;
{\cal T}(t',t){\cal Q}(t){\cal L}_{\rm SB}\sigma_\alpha\rangle^{(t')}
\quad.
\end{equation}
The GBE are non-Markovian:
our formalism allows for the inclusion of arbitrarily large memory effects.
Indeed, such non-Markovian effects will prove essential for the
determination of the correct relaxation rates.
\setcounter{equation}{0}
\section{Driven dynamics at weak dissipation}
For simplicity, we will from
now on set $\hbar=k_B=1$.
In order to evaluate the memory kernel we 
assume weak
coupling between the TLS and the bath,
and hence do lowest (i. e., second) order perturbation theory
in $H_{\rm SB}$ (Born approximation).
To this order we can omit the complement
projectors ${\cal Q}$, replace the evolution operator ${\cal T}$
with
\begin{equation}
{\cal T}(t',t)\,\to\,{\cal U}_S(t',t)
\otimes {\cal U}_B(t',t)
\quad,
\end{equation}
where ${\cal T}$, ${\cal U}_{\rm S}$,
${\cal U}_{\rm B}$ are the evolution operators
associated with ${\cal QLQ}$,
${\cal L}_{\rm S}$ and ${\cal L}_{\rm B}$, respectively,
and evaluate the scalar product $\langle\cdot;\cdot\rangle^{(t')}$
in the state
\begin{eqnarray}
\rho_{\rm rel}^{(0)}(t)
&=&
\rho_{\rm B}\otimes \rho_{\rm S}[\vec{\lambda}(t)]
\nonumber \\
&:=&
{1\over Z_{\rm B}}\exp(-\beta H_{\rm B})
\otimes {1\over Z_{\rm S}}\exp(
\textstyle{1\over2}\beta\vec{\lambda}(t)\cdot\vec{\sigma})
\end{eqnarray}
rather than in $\rho_{\rm rel}(t)$.
We thus obtain
\begin{equation}\label{kernel_1}
K_{\alpha\beta}(t,t')
=
g(t-t')\cdot \Xi_{\alpha\beta}(t,t')
\end{equation}
with the bath relaxation function
\begin{equation}
g(t-t'):=
2\beta
\langle e;{\cal U}_{\rm B}(t',t)e\rangle_{\rm B}
=2\int_0^\infty d\omega\,{J(\omega)\over\omega}\cos(\omega(t-t'))
\end{equation}
and spin relaxation function
\begin{equation}
\Xi_{\alpha\beta}(t,t') :=
{\textstyle {1\over4}}
\langle [\sigma_z,\sigma_\beta]; {\cal U}_{\rm S}(t',t)
[\sigma_z,\sigma_\alpha]\rangle^{(t')}_{\lambda}
\quad.
\end{equation}
Here 
$\langle\cdot;\cdot\rangle_{\lambda}^{(t')}$ and
$\langle\cdot;\cdot\rangle_{\rm B}$ denote scalar products
evaluated in the states 
$\rho_{\rm S}[\vec{\lambda}(t')]$ and $\rho_{\rm B}$,
respectively.
As the spectral density $J(\omega)$ has a characteristic width
$\omega_c$, the bath relaxation function $g(\tau)$
--which is essentially its Fourier transform--
decays on a typical scale (``memory time'')
$\tau_{\rm mem}\sim 1/\omega_c$.
For super-Ohmic damping this memory time is the only characteristic
time scale of the bath, independent of the temperature.
This distinguishes the super-Ohmic from the Ohmic case
where there exists a second characteristic scale of order $(K T)^{-1}$,
where $K$ is the Kondo parameter.

Using now, in Born approximation,
\begin{equation}
\vec{p}(t')=\chi(t')\vec{\lambda}(t')
\end{equation}
with
\begin{equation}
\chi(t'):={1\over |\vec{\lambda}(t')|}\tanh{\beta |\vec{\lambda}(t')|\over 2}
\quad,
\end{equation}
and defining the
instantaneous equilibrium polarization
\begin{equation}
\langle\vec{\sigma}\rangle_{h(t')}:=
\chi(t')\vec{h}(t')
\quad,
\end{equation}
the GBE can be cast into the compact form
\begin{equation} \label{GBE1}
\dot p_\alpha(t)=
[\vec{p}(t)\times\vec{h}(t)]_\alpha
-\int_0^t dt'\, g(t-t')\sum_\beta\Xi_{\alpha\beta}(t,t')
\chi^{-1}(t')[p_\beta(t')-\langle\sigma_\beta\rangle_{h(t')}]
\quad.
\end{equation}
This approximate equation of motion, valid at weak dissipation, 
is still non-Markovian, and still nonlinear in $\vec p$.
Far from equilibrium the nonlinearities may become significant
and lead to a non-exponential relaxation of the polarization vector.
Also the external driving field enters nonlinearly,
through the time evolution operator in the spin relaxation function.

\setcounter{equation}{0}
\section{Linear response regime}\label{linear_response}

To test our generalized Bloch equations we show that in the linear
response regime they yield results consistent with earlier calculations.
We assume that the dynamics take place in the linear regime,
i. e., that at all times the polarization be close to its
instantaneous equilibrium, $\vec p \approx \langle{\vec\sigma}\rangle$.
In this regime we may evaluate both 
$\Xi(t,t')$ and $\chi^{-1}(t')$ in
the instantaneous equilibrium state
$\rho_{\rm S}[\vec{h}(t')]$,
rather than in the state
$\rho_{\rm S}[\vec{\lambda}(t')]$,
thus rendering the GBE linear in $\vec p$.
Approximating the spin relaxation function by
\begin{equation}\label{Xi0}
 \Xi^{(0)}(t,t')=
\left(\begin{array}{ccc}
\cos(\epsilon_0\tau) & - \bar{u}_0
\sin(\epsilon_0\tau) & 0 \\
 \bar{u}_0\sin(\epsilon_0\tau) & 
 u^2_0  + \bar{u}_0^2\cos(\epsilon_0\tau) & 0 \\
0 & 0 & 0
\end{array}
\right)
\end{equation}
with $\tau := t-t'$,
$\epsilon_0 := \sqrt{\Delta^2 + \Delta_0^2}$,
$u_0:=\Delta_0/\epsilon_0$ and $\bar{u}_0:=\Delta/\epsilon_0$,
and expanding the susceptibility 
in a Fourier series
 \begin{equation}\label{N14b}
\chi^{-1}(t) =
\sum_{m=-\infty}^\infty \chi_m^{-1}\, e^{-im\omega_L t}
\quad,
\end{equation}
the GBE can be solved by Laplace transformation. 
Keeping 
terms up to linear order in $\Omega$ one finds, after tedious
but straightforward calculations,
\begin{equation}
p_z(t) = p_z^{(0)}(t) + p_z^{(1)}(t)
\end{equation}
where the first term describes the transient 
dynamics, while the second term describes the steady-state dynamics.
 The transient term is given by 
the standard second order result  for a static bias,\cite{rem}
\begin{equation}\label{dyn0}
p_z^{(0)}(t) =   u_0^2  
   \,e^{-t/\tau_2^{(0)}}\,\cos(\epsilon_0 t-\varphi)/\cos\varphi
+  (\bar{u}_0^2  - p_z^{\rm eq})\,e^{-t/\tau_1^{(0)}} \ + \ 
p_z^{\rm eq}
\quad,
\end{equation}
with 
 $p_z^{\rm eq} = - \bar{u}_0 \tanh(\beta\epsilon_0/2)$
and 
$\cot\varphi = \epsilon_0\tau_2^{(0)}$,
and relaxation rates 
\begin{equation}\label{relax_rates}
(\tau_2^{(0)})^{-1} =   (2\tau_1^{(0)})^{-1} =
   {\textstyle{1\over 2}} u_0^2\pi U \epsilon_0^3\coth(\beta\epsilon_0/2)
\quad.
\end{equation}
The steady-state term gives the linear response of the TLS to the driving field. 
In the asymptotic regime where all transients have decayed,
the polarization oscillates coherently with the driving frequency $\omega_L$;
in the linear regime there is no generation of higher harmonics. 
Defining
 the  dynamical susceptibility $\chi (\omega_L)$ via
the relation
\begin{equation}\label{VL2}
p_z^{(1)}(t) = - 2 \Omega \,\Big(\,\chi (\omega_L) 
                             \,e^{-i\omega_L t}
                               + \chi (-\omega_L)
                                \,e^{i\omega_L t}\,\Big)
\end{equation} 
we find for resonant driving ($\omega_L\approx \pm\epsilon_0$)
\begin{equation}\label{VL4}
\chi_{\rm res} (\omega_L) = {\textstyle{u_0^2 \over 4}}\, \tanh(\beta\epsilon_0/2) \sum_\pm 
\frac{\pm  i\tau_2^{(0)}}{1-i(\omega_L\mp\epsilon_0)\,\tau_2^{(0)}}
\end{equation} 
and for low-frequency driving ($\omega_L\ll \epsilon_0$)
\begin{equation}\label{VL6}
\chi_{\rm rel} (\omega_L)  - \chi_{\infty}\,=\,  
              \frac{\beta/4}{\cosh^2(\beta\epsilon_0/2)} 
            \frac{\bar{u}_0^2}{1-i\omega_L\tau_1^{(0)}}
\quad,
\end{equation}
where $\chi_\infty = (u_0^2/4\epsilon_0) \tanh(\beta\epsilon_0/2)$
denotes the instantaneous response. 
In the derivation we have made use of
$\chi_0\chi_{\pm 1}^{-1}  = \frac{\Omega\bar{u}_0}{\epsilon_0} 
[ 1 - (\beta\epsilon_0 
\coth(\beta\epsilon_0/2))/(2\cosh^2(\beta\epsilon_0/2))]$. 
These results are well-known from the literature, see e. g.
Ref. \CITE{HA}.

\setcounter{equation}{0}
\section{Effect of weak dissipation on dynamical localization}
%
%
We return to the general case of dynamics arbitrarily far from
equilibrium.
We  factor out that part of the dynamics
which is due to the driving field alone,
by defining new polarization vectors
\begin{equation}
\widetilde{p}_\alpha(t):=\sum_\beta R_{\alpha\beta}(t)p_\beta(t)
\quad,
\end{equation}
\begin{equation}
\langle\widetilde{\sigma}_\alpha\rangle_{h(t)}:=
\sum_\beta R_{\alpha\beta}(t)\langle\sigma_\beta\rangle_{h(t)}
\end{equation}
via an orthogonal rotation matrix
\begin{equation}
R(t):=\left(
\begin{array}{ccc}
\cos F(t) & \sin F(t) & 0 \\
-\sin F(t) & \cos F(t) & 0 \\
0 & 0 & 1
\end{array}
\right)
\end{equation}
with
\begin{equation}\label{F}
F(t):= \int_0^t f(s) ds = {2\Omega\over\omega_L}\sin(\omega_L t)
\quad.
\end{equation}
This rotation affects only the $x$- and $y$-components of
$\vec{p}(t)$, but not $p_z(t)$.
The thus rotated polarization vectors obey an equation of
motion
in which the driving field no longer appears explicitly, but
only indirectly via $R(t)$;
the only field which still appears explicitly is the static part
\begin{equation}
\vec{h}_{\rm stat}:=\overline{\vec{h}(t)}^t
=(\Delta_0,0,-\Delta)
\quad.
\end{equation}
The spin relaxation function in the GBE must be rotated, too,
and is now given by
\begin{equation}\label{rotated_relax}
(R(t)\Xi (t,t')R^{-1}(t'))_{\alpha\beta}={\textstyle {1\over4}}
\langle [\sigma_z,\sigma_\beta]; 
{\cal U}^{\Omega}(0,t'){\cal U}_{\rm S}(t',t){\cal U}^{\Omega}(t,0)
[\sigma_z,\sigma_\alpha]\rangle_\lambda^{(t')}
\quad,
\end{equation}
where ${\cal U}^{\Omega}$ denotes the spin evolution operator associated
with the driving field alone, at zero static field.
Since
\begin{equation}
(R(t)\Xi (t,t')R^{-1}(t'))_{zi} = 0
\quad\forall\, i=x,y,z
\quad,
\end{equation}
the $z$-component of the polarization vector satisfies
the equation of motion
\begin{equation}
{d\over dt}\widetilde{p}_z(t) = -\Delta_0 
\, [ \cos F(t)  \widetilde{p}_y(t) + \sin  F(t)  \widetilde{p}_x(t) ]
\quad.
\end{equation}

This
equation, together with the 
corresponding (more complicated)
equations for $\widetilde{p}_x(t)$ and $\widetilde{p}_y(t)$, has
provided the basis for an earlier
analysis \cite{DM} of the unbiased ($\Delta = 0$),
dissipationless TLS.
In Ref. \CITE{DM} the authors find
in the {\em fast-driving limit} ($\omega_L\gg \Delta_0$)
the driven  TLS to be equivalent to a
TLS without external field, but with renormalized 
 tunneling parameter $\Delta_0 J_0(2\Omega/\omega_L)$
instead of $\Delta_0$; and,
accordingly,  describe the dynamics by 
\begin{equation}
\widetilde{p}_z(t) = \cos [\Delta_0 J_0(2\Omega/\omega_L)\,t]
\quad.
\end{equation}
The authors conclude that in the fast-driving limit
and for small but nonzero $J_0(2\Omega/\omega_L)$ 
the dynamics exhibits strong slow mode oscillations
(``low-frequency generation''); while for
$J_0(2\Omega/\omega_L) = 0$ the component $\widetilde{p}_z(t)$
stops evolving at all: it remains forever ``frozen''
in its initial state
(``dynamical localization''). 

In order to  understand and later generalize
this result in our formalism
we, too, invoke the fast-driving limit:
we assume that the driving frequency be much
larger than both tunneling frequency and  bias,
\begin{equation}
\omega_L\gg\Delta_0, \Delta\ .
\end{equation}
Yet we do not assume the TLS to be unbiased or dissipationless.
The fast-driving limit 
 allows one to replace 
rapidly oscillating terms by their time averages:
\begin{eqnarray}\label{renpara}
R(t)\vec{h}_{\rm stat} &\to& 
\overline{ R(t)\vec{h}_{\rm stat}}^{t}
=(\Delta_0 J_0(2\Omega/\omega_L),0,-\Delta)
=: \vec{h}_{\rm eff}\\
\langle\vec{\widetilde\sigma}\rangle_{h(t')} &\to&
\overline{ \chi(t') R(t') \vec{h}(t') }^{t'}
=: \vec{p}_{\rm as}\\
(R(t)\Xi(t,t') R^{-1}(t'))_{\alpha\beta} 
&\to& 
\overline{(R(t)\Xi(t,t') R^{-1}(t'))}^{t'}_{\alpha\beta}
=: \widetilde{\Xi}_{\alpha\beta}(t-t')
\quad. \label{MM}
\end{eqnarray}
The latter average is taken at fixed $(t-t')$.
With these replacements the GBE (\ref{GBE1}) acquires the convolution form
\begin{equation} \label{GBE2}
\dot{\widetilde{p}}_\alpha (t) = 
[\vec{\widetilde{p}}(t)\times \vec{h}_{\rm eff}]_\alpha
- \int_0^t dt'\,g(t-t')
\sum_\beta \widetilde{\Xi}_{\alpha\beta}(t-t')\,
\chi^{-1}(t')[\widetilde{p}_\beta(t') - p^{\rm as}_\beta]
\quad;
\end{equation}
in particular,
\begin{equation}\label{cc}
{d\over dt}\widetilde{p}_z(t) 
= - J_0(2\Omega/\omega_L)\,\Delta_0\, \widetilde{p}_y(t) 
\end{equation}
which predicts a slowing down of the time evolution 
of $\widetilde{p}_z(t)$ near $J_0 = 0$.

To this time-averaged GBE there are rapidly oscillating
correction terms which are negligible as long as
$J_0\ne 0$, but which may become important at the
localization transition $J_0=0$.
Indeed,
numerical studies by Makarov and Makri\cite{Maki} indicate
that at $J_0=0$ the rapidly oscillating correction terms 
cause additional phase relaxation
(if the system is coupled to a heat bath),
and eventually destroy the dynamical localization.
Hence the time-averaged GBE (\ref{GBE2}) can only describe the
dynamics near, but not exactly at, the localization transition.
Furthermore, even away from the localization transition,
the time-averaged GBE
  misses short time transients (Gaussian or algebraic decay) 
arising from 
the bath decorrelation and from  the nonlinear 
time evolution induced by the laser 
field.
Our  description is thus expected to be 
valid on intermediate time scales 
$\omega_c^{-1}, \omega_L^{-1}\ll t \ll \tau_\alpha$, 
where the $\{\tau_\alpha^{-1}\}$ are the rates at which
the various components of the polarization vector $\vec{\widetilde{p}}(t)$
relax 
to their stationary state.

A  full analytical solution of the dynamics of the polarization vector
$\vec{\widetilde p}(t)$, and hence a study of its dynamics close to
the roots of the Bessel function $J_0$,
 is only possible if we invoke  further approximations:
namely, the {\it high-temperature limit},  $T\gg \Delta_0, \Delta, \Omega$,
and the {\it high-frequency limit}, $\omega_L\gg\omega_c$.
In the high-temperature limit
we may evaluate both 
$\Xi(t,t')$ and $\chi^{-1}(t')$ in
$\rho_{\rm S}[0]={\textstyle{1\over2}}{\bf 1}_{\rm S}$
rather than in the time-dependent state
$\rho_{\rm S}[\vec{\lambda}(t')]$,
allowing us to replace
\begin{equation}
\langle\cdot;\cdot\rangle_\lambda^{(t')}
\,\to\,
\langle\cdot;\cdot\rangle_0
\quad,\quad
\chi^{-1}(t')\to 2/\beta
\quad.
\end{equation}
The product of evolution operators 
in the rotated spin relaxation function (\ref{rotated_relax})
can be written as
a time-ordered exponential
\begin{equation} \label{T}
{\cal U}^{\Omega}(0,t'){\cal U}_{\rm S}(t',t){\cal U}^{\Omega}(t,0)
= {\rm T} \exp\left[-{ i\over2}\int_{t'}^t dt'' 
(R(t'') \vec{h}_{\rm stat})\cdot \vec{\sigma}^\times\right]
\quad,
\end{equation}
where $\sigma^\times$ denotes the commutator with $\sigma$.
In the high-frequency limit  the integrand in the exponent
oscillates rapidly during the interval $[t',t]$, which has a
typical length $\tau_{\rm mem}\sim1/\omega_c \gg 1/\omega_L$.
In order to perform the average  (\ref{MM}) we 
may thus replace
\begin{equation}
\int_{t'}^t dt'' (R(t'') \vec{h}_{\rm stat}) 
\longrightarrow (t-t') \cdot
\overline{ R(t'')\vec{h}_{\rm stat}}^{t''}
\quad,
\end{equation}
which will yield just
the time evolution operator 
of a static tunneling problem with new parameters (\ref{renpara}).
Defining
\begin{eqnarray}\label{dd1}
\epsilon&:=&\sqrt{\Delta^2+\Delta_0^2 J_0^2(2\Omega/\omega_L)}
\quad,
\\
u &:=&   \Delta_0 J_0(2\Omega/\omega_L)/\epsilon
\quad,
\\
\bar{u} &:=&  \Delta/\epsilon = \sqrt{1 -  u^2} 
\quad,\label{dd3}
\end{eqnarray}
the rotated spin relaxation function is then given by \begin{equation}
\widetilde{\Xi}(\tau)=
\left(\begin{array}{ccc}
\cos(\epsilon\tau) & -\bar{u}
\sin(\epsilon\tau) & 0 \\
\bar{u}\sin(\epsilon\tau) & 
 u^2  + \bar{u}^2\cos(\epsilon\tau) & 0 \\
0 & 0 & 0
\end{array}
\right)
\quad.
\end{equation}
 
We recognize that now
the dynamics has the same form as in the linear response regime
(Sec. IV), with the additional simplification
$\beta\epsilon\ll 1$.
Hence, indeed,  
the driven system can be mapped
onto a time-independent TLS with modified parameters.
The results for the time evolution of 
$\widetilde{p}_z(t)$ can immediately be
taken over from (\ref{dyn0}), with the sole replacement
\begin{equation}\label{re}
\Delta_0 \to \Delta_0 J_0(2\Omega/\omega_L)\ ;
\end{equation}
and consequently, 
$u_0 \to u$, $\bar{u}_0 \to \bar{u}$ and $\epsilon_0 \to \epsilon$.
Noting that 
$(\tau_\alpha^{(0)})^{-1} \propto 
\Delta_0^2$ (Eq. \ref{relax_rates})
we find that the relaxation rates 
must be modified according to
\begin{equation}
(\tau_\alpha^{(0)})^{-1} 
\to 
[J_0(2\Omega/\omega_L)]^2\,(\tau_\alpha^{(0)})^{-1}
\quad .
\end{equation} 

In the unbiased case ($\Delta = 0$) and 
away from the roots of the Bessel function  the system
thus reaches its steady state  on the time scale
$\tau_2^{(0)}/[J_0(2\Omega/\omega_L)]^2$. 
Near the localization transition ($J_0\to 0$)
this time scale diverges faster than 
the tunneling time $(\Delta_0 J_0(2\Omega/\omega_L))^{-1}$,
dissipation thus being effectively  switched off.
As one reads off from the linear response solution
(\ref{dyn0}), this implies that 
there will be
a slow mode coherent oscillation near the localization transition,
and that the particle will remain trapped in one well on the
time scale $(\Delta_0 J_0(2\Omega/\omega_L))^{-1}$.
This is consistent with
Dakhnovskii's earlier analysis  within the 
framework of the NIBA\cite{Yu1}, and with the ab initio 
numerical calculations of Makarov and Makri\cite{Maki}.

In the biased case ($\Delta\neq 0$), on the other hand,
the amplitude of the oscillatory term in (\ref{dyn0}),
\begin{equation}
u^2 = (J_0(2\Omega/\omega_L) \Delta_0)^2/\epsilon^2
\quad,
\end{equation}
becomes
negligible for $J_0(2\Omega/\omega_L)\ll \Delta/\Delta_0$.  
One thus expects pure exponential decay,
rather than slow mode oscillations,
in the immediate vicinity of $J_0 = 0$.
(One way of viewing this phenomenon is that as the bias
is switched on, the quasienergy levels cease to cross and
hence the destruction of tunneling can no longer be coherent.)
The particle will then remain trapped in one well
on the time scale $\tau_1^{(0)}/[J_0(2\Omega/\omega_L)]^2$. 
Further away from the localization transition the exponential decay will
be superimposed with small-amplitude
oscillations.

The evolution of $\widetilde{p}_z(t)$ in the two cases
(unbiased and biased) is
illustrated in Figs. 1 and 2, respectively.
We have calculated the curves using
Eq. (\ref{dyn0}), with the replacement (\ref{re}).

 \setcounter{equation}{0}
 \section{Summary and final remarks}
In this paper we have studied the influence of a weakly coupled 
 thermal environment on dynamical localization.
Motivated by the breakdown of the NIBA
at non-zero bias and weak dissipation 
we have investigated the dynamics of a biased two-level tunneling system
strongly driven by an external field
and weakly coupled to a super-Ohmic heat bath.
We have employed the Robertson projection operator formalism 
to derive generalized Bloch equations for the polarization
vector $\vec{p}(t)$.

Assuming only the validity of
the two-state picture and of the Born approximation,
and invoking the
fast-driving, high-frequency and high-temperature limits,
we have shown that 
on intermediate time scales the driven dissipative TLS 
can be mapped onto a dissipative TLS without driving,
but with renormalized parameters.
We have found that since  the phase 
relaxation rate decreases quadratically with 
$J_0(2\Omega/\omega_L)$,  while
the tunneling frequency decreases only linearly,
coherence
is restored
near the localization point of an  unbiased TLS ($\Delta = 0$).
This manifests itself in
slow mode coherent oscillations
near the localization transition, a result that
Dakhnovskii had
previously obtained within the framework of the NIBA.
In the biased case ($\Delta\ne 0$)
we found a qualitatively different
behavior, namely pure exponential decay in the immediate vicinity
of $J_0(2\Omega/\omega_L) = 0$.

Let us finally comment on two other papers 
 which address the problem of driven tunneling dynamics
in the weak-dissipation  limit.\cite{Milena,Ditt}
Dittrich, H\"anggi and Oelschl\"ager \cite{Ditt}
consider dynamical localization in the weak-coupling limit
(Born  approximation) and solve numerically
the quantum master equation 
for the full continuous bistable system with  Ohmic dissipation.
In contrast to our present analysis, however, they employ
a restricted
rotating-wave approximation and assume
the bath to be Markovian, i.~e., they assume that the correlation time
for the boson modes is negligible compared to the
characteristic time scale  of the double-well dynamics.
For Ohmic friction  this assumption is justified if 
in addition to the classical time scale $\tau_{\rm mem}
\sim 1/\omega_c$ also the quantum time scale 
of the bath $(KT)^{-1}$, with $K$ being the 
Kondo parameter, is shorter
 than the tunneling time.
But in this case the dynamics is incoherent
and a weak-coupling picture  inappropriate.
In fact, it has been shown explicitly
that the Born approximation 
fails for a tunneling system
subject to Ohmic dissipation {\it and} driving, and that
driving renders the dynamics intrinsically non-Markovian.\cite{Ma}
This is not surprising since even for super-Ohmic dissipation,
where the second (quantum) time scale does not
exist,  non-Markovian effects are essential for the determination
of the correct relaxation rates.\cite{sohm}

Very recently Grifoni {\it et al.}\cite{Milena}
have extended their previous treatment\cite{Ma}
{beyond the NIBA}.  They derive an exact master equation,
which they then use to  determine the dynamics 
at weak coupling and fast driving.
Indeed, our GBE  finds a counterpart 
in the set of equations adjacent to their Eq. (10).
Yet instead of discussing  the influence of weak dissipation
on the dynamics near the 
localization transition, they use their master equation
as a starting point to investigate the modification
of quantum coherence {\em away} from the roots of the Bessel
functions, in the regime $\Omega/\omega_L \ll 1$.

\acknowledgements 
          We  wish to thank  R. J. Silbey,   M. Cho, and 
M. Grifoni  for  helpful discussions. Financial support 
by  NSF and the   Alexander von  Humboldt foundation (P.N.), as well
as by the European Union HCM programme (J.R.),
is gratefully acknowledged.



\newpage

\section*{Figure Captions}

\begin{itemize}
\item[FIG. 1] Time evolution of $\widetilde{p}_z(t)$ in the symmetric case.
The parameters are
$T = 10 \Delta_0$, $\Delta_0 \tau_2^{(0)} = 10$ and $\Delta = 0$,
for driving fields with
$2\Omega/\omega_L = 0$
(--$\cdot$--),  $2\Omega/\omega_L = 2$
(--  --), and   $2\Omega/\omega_L = 2.3$  (---). 
The localization transition
occurs at $2\Omega/\omega_L \approx 2.405$. 
\item[FIG. 2] Time evolution of $\widetilde{p}_z(t)$ in the biased case.
The parameters are chosen as in Fig. 1, except for
$\Delta = 0.75 \Delta_0$.
\end{itemize}


\begin{references}
\bibitem{footnote_born}
     The Born approximation assumes not only that dissipation is
sufficiently weak, but also that the phonon density of states falls off
rapidly enough at low energies.
This latter condition is satisfied for, e. g., super-Ohmic
damping or a 3D-Debye spectrum.
For Ohmic damping, however, the Born approximation is generally very poor
and actually breaks down in the presence of a driving field.\cite{Ma}
In such cases a strong coupling approach like the NIBA is more
appropriate.
\bibitem{rem} Here we neglect the  frequency shift
              induced by the bath. This will not 
              alter any of our conclusions.
\bibitem{sohm} Even in the linear response regime, the
               familiar expression (\ref{relax_rates})
               cannot be derived in the Markov limit of the GBE.
               Instead a pole-approximation,
               or, equivalently, a 
              formulation of the dynamics in the interaction picture
              has to be used.
\end{references}
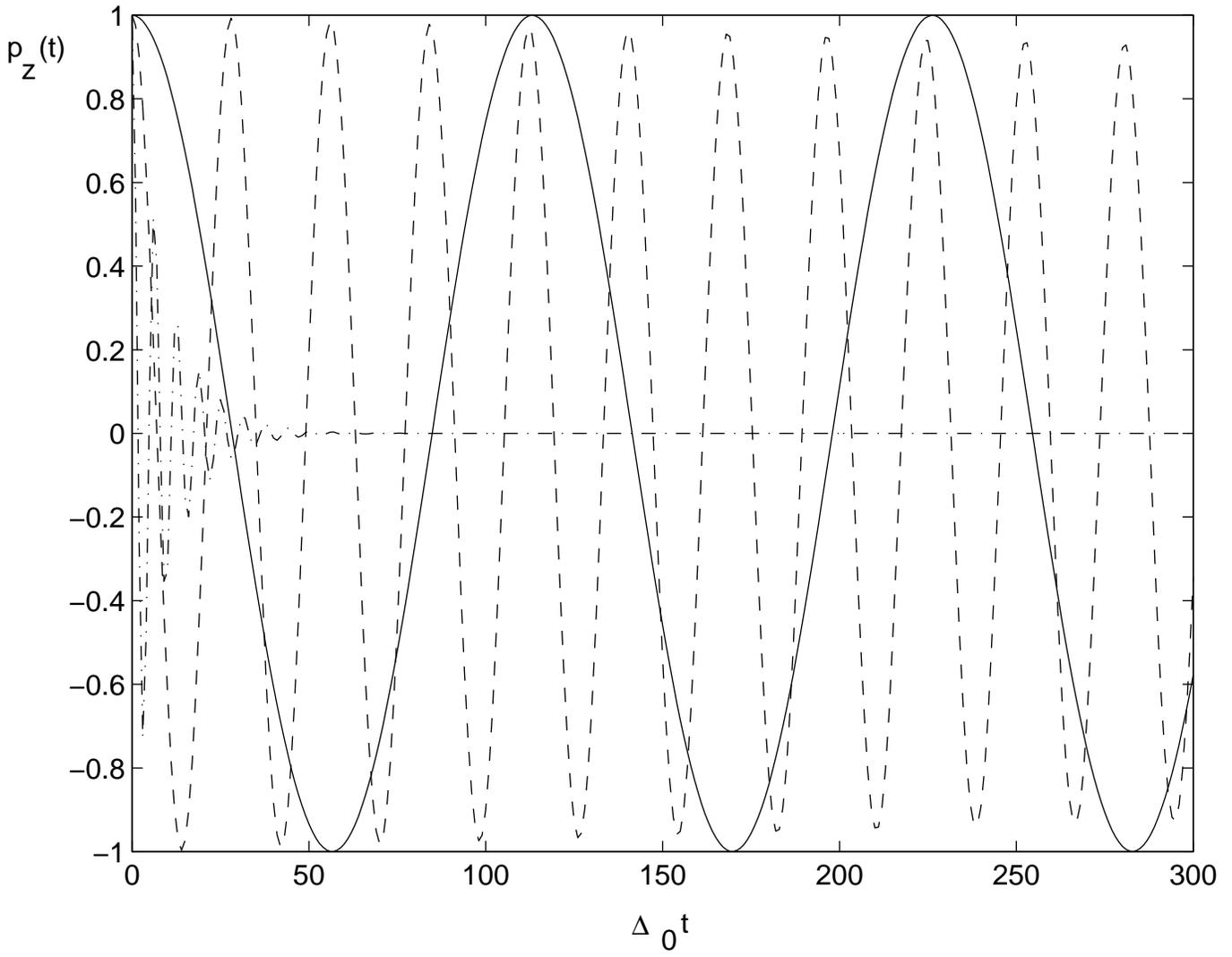
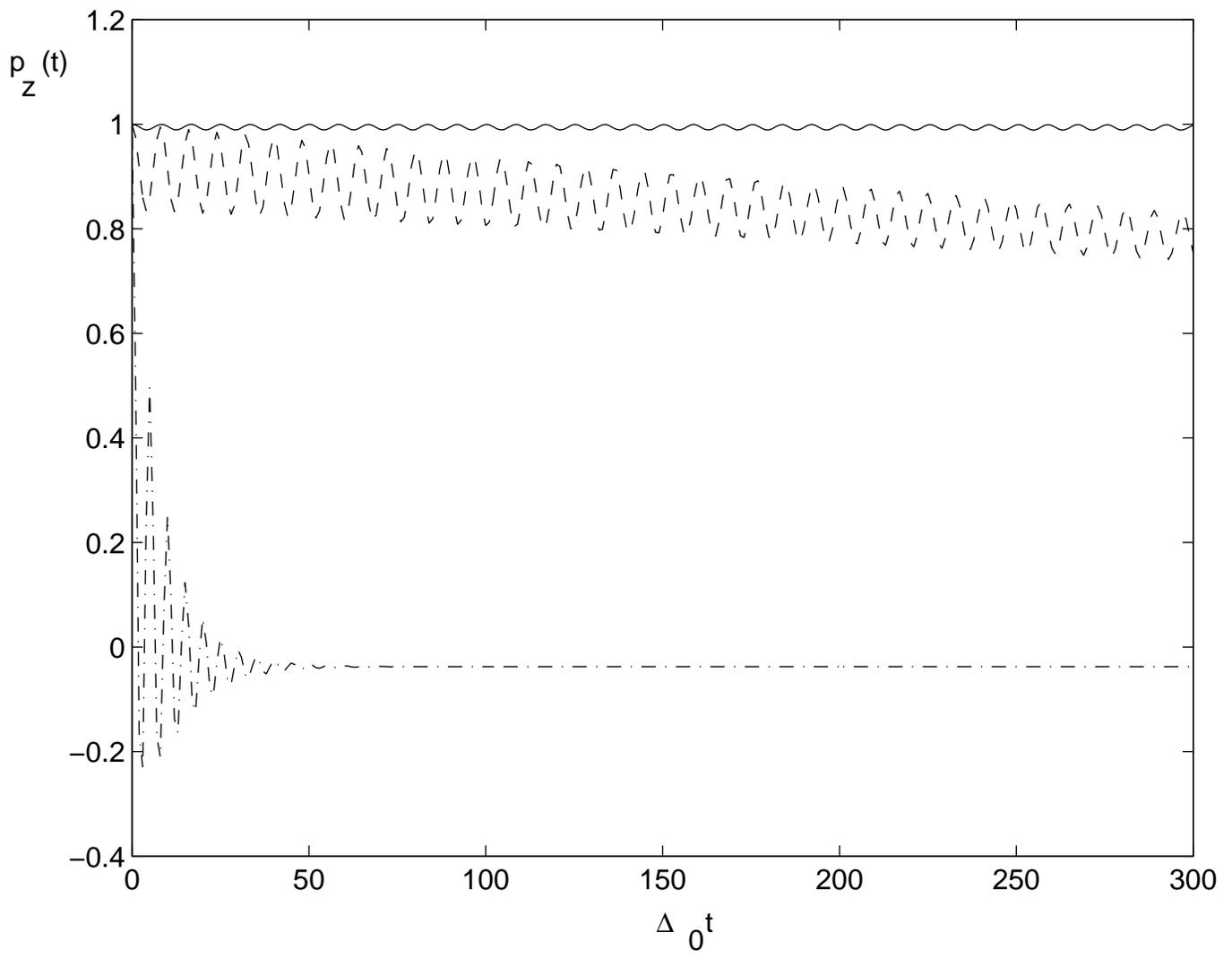
\end{document}